\pdfoutput=1
\documentclass[a4paper, times, 12pt]{cas-sc}

\usepackage[numbers,sort&compress]{natbib}
\usepackage{color}
\usepackage{caption}
\usepackage{float}
\usepackage{diagbox}
\usepackage{url}
\usepackage{mathtools,amssymb}
\usepackage{lineno,hyperref}
\usepackage[numbers]{natbib}

\usepackage{algorithm}
\usepackage{algorithmicx}
\usepackage{algpseudocode}
\usepackage{subfigure}

\modulolinenumbers[5]
\usepackage{verbatim}
\usepackage{amsmath}
\usepackage{algorithm}
\usepackage{algpseudocode}
\usepackage{caption}
\usepackage{amsmath}
\usepackage{amssymb}
\usepackage{siunitx}
\usepackage{pifont}
\usepackage{colortbl}
\definecolor{lightblue}{RGB}{220,230,241} 
\def\tsc#1{\csdef{#1}{\textsc{\lowercase{#1}}\xspace}}
\tsc{WGM}
\tsc{QE}
\tsc{EP}
\tsc{PMS}
\tsc{BEC}
\tsc{DE}
\begin{document}
\let\WriteBookmarks\relax
\def\floatpagepagefraction{1}
\def\textpagefraction{.001}
\shorttitle{}
\shortauthors{Haishan Huang \textit{et al.}}
\captionsetup[figure]{labelfont={bf},labelsep={period},name={Fig.}}

% \title [mode = title]{TopoWMamba: Topology-Aware Wavelet Mamba for Airway Structure Segmentation in Postoperative Recurrent Nasopharyngeal Carcinoma CT Scans} 
\title [mode = title]{Topology-Aware Wavelet Mamba for Airway Structure Segmentation in Postoperative Recurrent Nasopharyngeal Carcinoma CT Scans} 

\author[secondaddress]{Haishan Huang\texorpdfstring{$^{\dagger}$}{}}
\author[mymainaddress]{Pengchen Liang\texorpdfstring{$^{\dagger}$}{}}
\author[secondaddress12]{Naier Lin\texorpdfstring{$^{\dagger}$}{}}
\author[secondaddress12]{Luxi Wang}
\author[six]{Bin Pu}
% 22211320010@m.fudan.edu.cn luqianwang
\author[secondaddress]{Jianguo Chen}
\ead{chenjg33@mail.sysu.edu.cn}
\cormark[1]
\author[fouraddress]{Qing Chang}
\ead{robie0510@hotmail.com}
\cormark[1]
\author[secondaddress11]{Xia Shen}
\ead{zlsx@yahoo.com}
\cormark[1]
\author[secondaddress11]{Guo Ran}
\ead{ranguo@eentanesthesia.com}
\cormark[1]

% \author[six]{Bin Pu}
% eebinpu@ust.hk

% Naier Lin
% linnaierbaby@qq.com

% Luxi Wang
% 22211320010@m.fudan.edu.cn

% \author[four]{Huiping Yao}
% \ead{yhp12680@rjh.com.cn}
% \cormark[1]
% \author[fouraddress]{Qing Chang}
% \ead{robie0510@hotmail.com}
% \cormark[1]

\address[secondaddress]{School of Software Engineering, Sun Yat-sen University, 519000, Zhuhai, Guangdong Province, China}
\address[mymainaddress]{School of Microelectronics, Shanghai University, 201800, Shanghai, China}
\address[six]{Electronic and Computer Engineering, The Hong Kong University of Science and Technology, 999077, Hong Kong, China}
% \address[five]{School of Public Administration, East China Normal University, 200062, Shanghai, China}
\address[fouraddress]{Department Shanghai Key Laboratory of Gastric Neoplasms, Department of Surgery, Shanghai Institute of Digestive Surgery, Ruijin Hospital, Shanghai Jiao Tong University School of Medicine, 200025, Shanghai, China}
\address[secondaddress12]{Department of Radiology, Eye \& ENT Hospital, Fudan University, Shanghai, 200031, China}
\address[secondaddress11]{Department of Anesthesiology, Eye \& ENT Hospital, Fudan University, Shanghai, 200031, China}
% \address[secondaddress11]{Department of Anesthesiology, Eye \& ENT Hospital, Fudan University, Shanghai, 200031, China}

% Footnotes for equal contribution
\fntext[fn1]{Haishan Huang, Pengchen Liang, and Naier Lin contributed equally to this work.}
\cortext[cor1]{Corresponding author}

\begin{abstract}
Nasopharyngeal carcinoma (NPC) patients often undergo radiotherapy and chemotherapy, which can lead to postoperative complications such as limited mouth opening and joint stiffness, particularly in recurrent cases that require re-surgery. These complications can affect airway function, making accurate postoperative airway risk assessment essential for managing patient care.
Accurate segmentation of airway-related structures in postoperative CT scans is crucial for assessing these risks. This study introduces TopoWMamba (Topology-aware Wavelet Mamba), a novel segmentation model specifically designed to address the challenges of postoperative airway risk evaluation in recurrent NPC patients. TopoWMamba combines wavelet-based multi-scale feature extraction, state-space sequence modeling, and topology-aware modules to segment airway-related structures in CT scans robustly. By leveraging the Wavelet-based Mamba Block (WMB) for hierarchical frequency decomposition and the Snake Conv VSS (SCVSS) module to preserve anatomical continuity, TopoWMamba effectively captures both fine-grained boundaries and global structural context, crucial for accurate segmentation in complex postoperative scenarios.
Through extensive testing on the NPCSegCT dataset, TopoWMamba achieves an average Dice score of 88.02\%, outperforming existing models such as UNet, Attention UNet, and SwinUNet. Additionally, TopoWMamba is tested on the SegRap 2023 Challenge dataset, where it shows a significant improvement in trachea segmentation with a Dice score of 95.26\%. The proposed model provides a strong foundation for automated segmentation, enabling more accurate postoperative airway risk evaluation.
\end{abstract}


\maketitle

\section{Introduction}

Nasopharyngeal carcinoma (NPC) is a common cancer endemic to Southeast Asia and Southern China, with complex treatment challenges due to its anatomical proximity to critical structures~\cite{wong2021nasopharyngeal}. 
While primary radiotherapy and chemotherapy achieve high initial control rates, recurrent cases often require salvage surgery, which carries significant postoperative risks, particularly airway-related complications such as stenosis and obstruction~\cite{peng2022treatment}. 
These complications, exacerbated by prior treatments-induced fibrosis and tissue remodeling, can severely compromise respiratory function and long-term survival~\cite{chua2016nasopharyngeal,kushihashi2024case,mou2011surgical}. 
Accurate assessment of postoperative airway risks is therefore critical for optimizing patient management.

Accurate segmentation of airway-related structures (e.g., pharynx, larynx, trachea) in postoperative CT images is essential for risk evaluation. However, postoperative CT scans present unique challenges, including tissue deformation, scar formation, and blurred boundaries between adjacent structures~\cite{islam2023deep}. Traditional segmentation methods, such as atlas-based approaches, fail to adapt to these morphological variations~\cite{wang2022evaluation}. Although deep learning models (e.g., UNet variants) have advanced general anatomical segmentation, they frequently overlook subtle postoperative alterations—such as radiation-induced fibrosis or surgical artifact distortions—due to their limited ability to model texture heterogeneity and global anatomical continuity~\cite{zhang2024weakly}. 

Recent advances in multi-scale feature learning offer promising solutions to these challenges. Wavelet transforms, which decompose images into frequency bands at multiple scales, enable the simultaneous analysis of both coarse anatomical shapes and detailed boundary structures~\cite{zhou2023xnet}. When paired with state-space modeling, which efficiently captures global contextual dependencies across CT slices, this approach addresses both tissue continuity and localized postoperative variations~\cite{feng2024wavelet}. 
Furthermore, topology-aware modules, such as Snake Conv VSS (SCVSS) \cite{zuo2024topology}, align feature extraction with the intrinsic geometry of structures, improving boundary detection in ambiguous regions.

While most segmentation methods focus on tumor detection or general anatomical structures, airway segmentation in postoperative CT scans requires models that can handle the complexity and subtlety of these structures~\cite{zhou2021review}. Deep learning techniques, including CNNs and transformer-based models, excel at general segmentation tasks~\cite{ronneberger2015u, chen2021transunet,yuan2023effective}, but often fall short when applied to the intricate anatomy of the airway and surrounding structures.

To address these limitations, we propose the TopoWMamba model, which integrates wavelet transforms for multi-scale feature extraction and state-space sequence modeling. This model is designed to accurately segment airway-related structures in CT scans of recurrent NPC patients, providing a solid foundation for future risk prediction models.

In this paper, we make the following key contributions:
\begin{itemize}
\item We propose the integration of wavelet transforms with state-space modeling, enabling the TopoWMamba model to effectively extract multi-scale features, significantly improving the segmentation of complex, anatomically varied airway structures in postoperative CT scans.
\item We introduce the Snake Conv VSS (SCVSS) module, which optimizes the detection of complex boundaries in airway-related structures, particularly in regions where postoperative changes might complicate segmentation. This module ensures that small and intricate structures, such as the larynx and pharynx, are more accurately segmented.
\item We establish the NPCSegCT dataset, a comprehensive collection of CT scans from recurrent NPC patients. The dataset includes detailed annotations of critical airway-related structures, providing a high-quality resource for training and evaluating segmentation models.
\item We demonstrate through rigorous testing on the NPCSegCT dataset that TopoWMamba outperforms existing models in terms of segmentation accuracy, with an average Dice score of 88.02\%. This performance highlights its potential to support future research on airway risk prediction models.
\end{itemize}

\begin{figure*}[htbp]
\centering
\includegraphics[width=\textwidth]{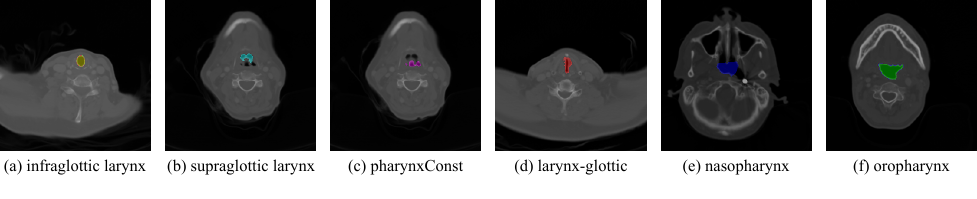}
\caption{Partial display of our NPCSegCT dataset, showcasing annotated CT scans with critical airway-related structures.}
\label{fig:fig1}
\end{figure*}

\section{Related Work}

\subsection{Deep Learning Approaches for Medical Image Segmentation}
Deep learning has fundamentally transformed medical image segmentation over the past decade \cite{wang2022medical}. 
Early architectures such as U-Net~\cite{ronneberger2015u} and its numerous variants~\cite{qamar2020variant,lin2024mm} employ an encoder-decoder structure with skip connections to effectively capture both contextual and spatial information~\cite{azad2024medical}. 
These models have achieved remarkable success in various segmentation tasks across different imaging modalities, including MRI~\cite{guo2022cardiac}, CT~\cite{yang2024novel}, and ultrasound~\cite{chen2023rrcnet}. 
More recently, transformer-based models like TransUNet~\cite{chen2021transunet} and Swin-UNet~\cite{cao2022swin} have further improved performance by integrating self-attention mechanisms to capture long-range dependencies, demonstrating superior results in tasks such as brain tumor segmentation~\cite{zhang2021transfuse} and organ segmentation. 
Despite these advances, many current approaches struggle with preserving fine anatomical boundaries and maintaining spatial coherence, particularly in challenging scenarios such as postoperative nasopharyngeal carcinoma imaging, where critical airway-related structures lie nearby. 
These limitations motivate the need for models that can more effectively capture both local details and global context, as highlighted in recent studies~\cite{zhang2022understanding,rayed2024deep}.

\subsection{Frequency Domain Analysis and Wavelet-based Methods in Medical Imaging}
Frequency domain analysis has long been a powerful tool in image processing, with wavelet transforms playing a central role in multi-scale feature extraction \cite{sun2021mwq}. 
Wavelet-based techniques excel at decomposing images into components that capture both local details (high-frequency components) and global structures (low-frequency components) \cite{liang2024rskd}. 
Recent research has integrated wavelet transforms into deep learning frameworks to bolster the extraction of robust features and enhance segmentation performance~\cite{tan2024wavelet,yang2024sffnet}. By leveraging both spatial and frequency domain information, such methods can better capture subtle textural variations and edge details~\cite{liu2024freqsnet}. 
The integration of wavelet transforms into frameworks allows for efficient processing of multi-scale features, thereby improving the reliability and accuracy of segmentation outputs~\cite{qian2024adaptive}. 
% Furthermore, wavelet-based approaches have been shown to outperform traditional methods across various imaging modalities, including MRI and CT~\cite{alijamaat2021multiple,agnes2024wavelet}. 
This trend emphasizes the potential of wavelet transforms when combined with machine learning techniques to enhance medical image analysis.

\subsection{State Space Sequence Models and Topology-aware Techniques}
State space sequence models (SSMs) have emerged as an attractive alternative to traditional attention mechanisms, particularly due to their ability to model long-range dependencies with linear computational complexity \cite{zhu2024vision}. 
The Mamba architecture~\cite{gu2023mamba} is a notable example, employing selective state space modeling to capture global contextual cues that are crucial for maintaining anatomical consistency across complex structures. 
% SSMs proposed by Gu et al.~\cite{gu2021efficiently} have shown great promise in various medical imaging tasks, leveraging their efficiency in handling sequential information.
In parallel, topology-aware techniques have been proposed to ensure that segmentation outputs preserve the natural spatial relationships among anatomical structures \cite{sadikine2024deep,puunsupervised}. 
Techniques such as topology-preserving segmentation~\cite{santhirasekaram2023topology,shi2023nextou} address common issues like fragmented or disconnected segmentations, which can lead to clinically unacceptable results. 
For instance, Gupta et al.~\cite{gupta2022learning} demonstrated how incorporating topological constraints can enhance the robustness of segmentation algorithms, particularly in challenging cases such as airway-related structure delineation.
In our work, we extend these ideas by introducing a topology-aware snake-scan module that adaptively reorders feature patches to enhance boundary delineation and preserve the inherent topology of airway-related structures.

\begin{figure*}[htbp]
\setlength{\abovecaptionskip}{2pt}
\setlength{\belowcaptionskip}{0pt}
\centering
\includegraphics[width=\textwidth]{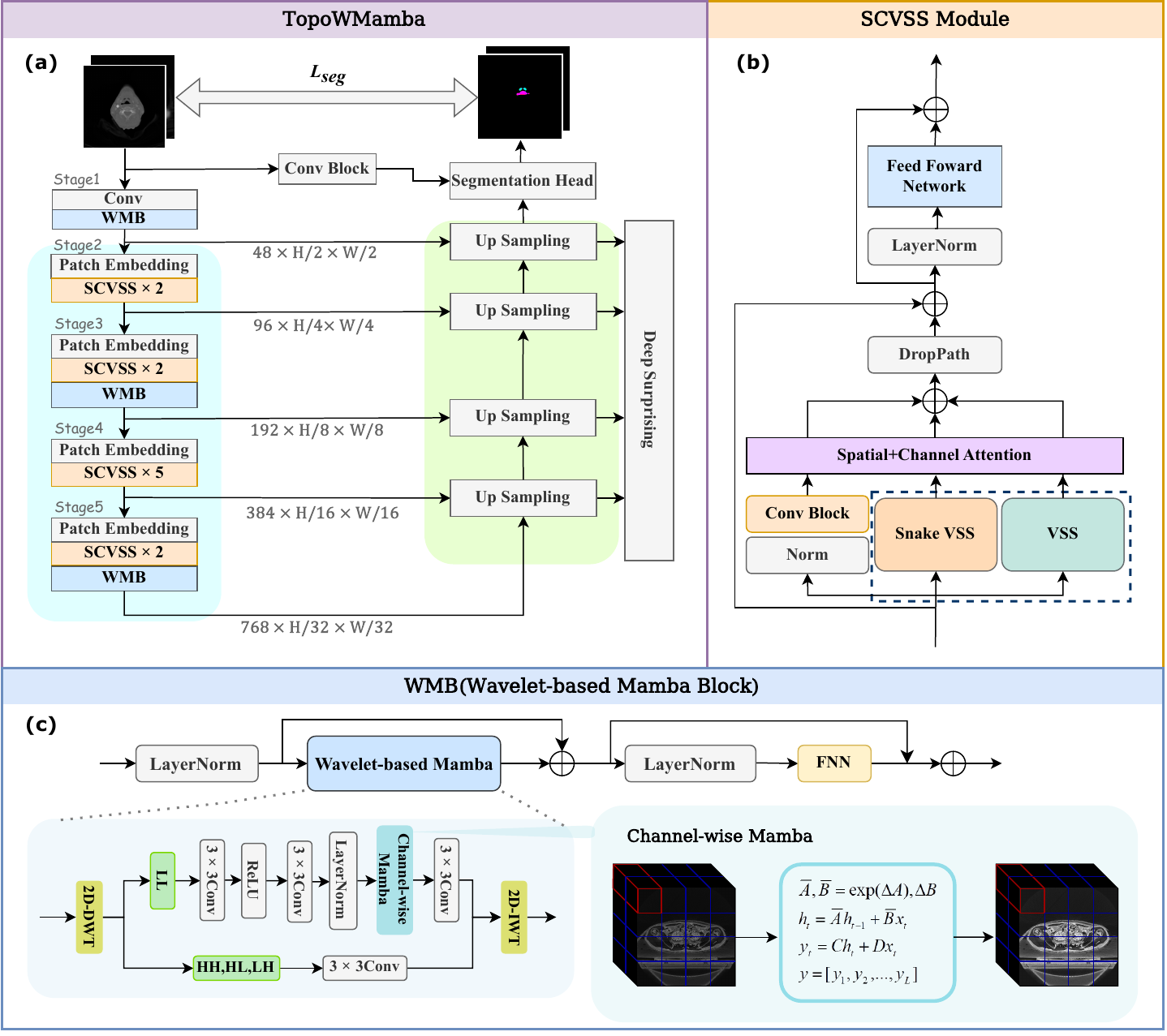}
\caption{(a)The architectural design of TopoWMamba. TopoWMamba is an encoder-decoder segmentation framework that employs Mamba-based modules for effective feature extraction while maintaining low-level details through residual connections. (b)The overall structure of the SCVSS. The SCVSS features three parallel branches—conventional convolution, VSS, and SnakeVSS. (c)The illustration of Wavelet-based Mamba Block (WMB). WMB utilizes a 2D discrete wavelet transform to separate feature maps into low and high-frequency components, processing them with specialized modules to enhance long-range dependencies and global context.}
\label{fig:fig2}
\end{figure*}

\begin{figure}[htbp]
\centering
\includegraphics[width=2.5in]{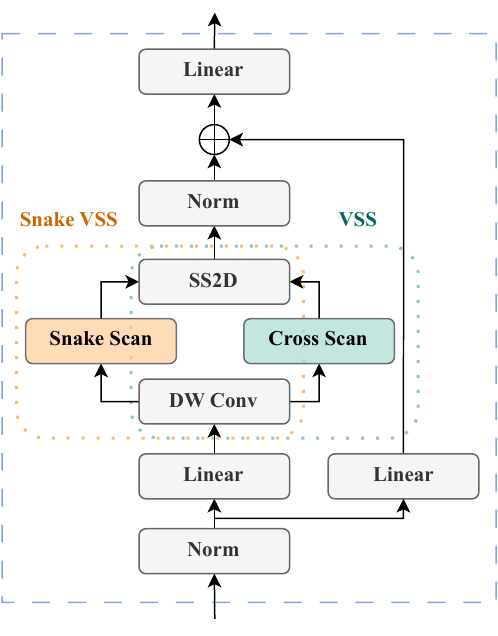}
\caption{Details of SnakeVSS and VSS structure. In this diagram, the symbol $\oplus$ represents element-wise addition. The SnakeVSS branch reorders feature patches in serpentine patterns, capturing complex curvilinear structures, while the VSS branch focuses on conventional scanning directions to extract spatial features effectively.}
\label{fig:fig4}
\end{figure}

% \begin{figure}[htbp]
% \centering
% \includegraphics[width=2.5in]{Figure5.pdf}
% \caption{Details of SnakeVSS and VSS structure. In this diagram, the symbol $\oplus$ represents element-wise addition. The SnakeVSS branch reorders feature patches in serpentine patterns, capturing complex curvilinear structures, while the VSS branch focuses on conventional scanning directions to extract spatial features effectively.}
% \label{fig:fig4}
% \end{figure}

\begin{figure}[htbp]
\setlength{\abovecaptionskip}{2pt}
\setlength{\belowcaptionskip}{0pt}
\centering
\includegraphics[width=0.5\textwidth]{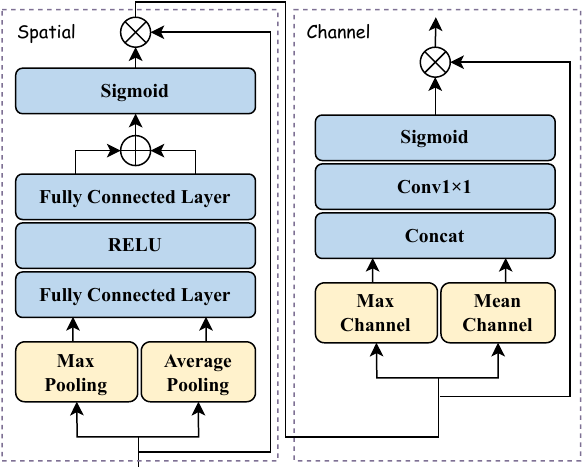}
\caption{Details of spatial and channel attention structure. The symbol $\otimes$ denotes element-wise multiplication, and $\oplus$ represents element-wise addition.  This structure enhances feature representation by focusing on important spatial regions and channel-wise dependencies, allowing the model to better capture relevant information.}
\label{fig:fig5}
\end{figure}

\section{Method}

We propose an efficient encoder–decoder segmentation framework that integrates Mamba-based modules to extract both global and local features while preserving low-level details through residual connections and deep supervision. The overall architecture consists of a Mamba-based encoder, a decoder with up-sampling blocks, and a segmentation head that fuses multi-scale features via skip connections. This design ensures high segmentation accuracy while maintaining computational efficiency, as illustrated in Fig.~\ref{fig:fig2}.

\subsection{Encoder}

The encoder is organized into five stages, each performing a $2\times$ down-sampling operation. In the first stage, a $7\times7$ convolution with stride $2$ and padding $3$ reduces the input image of size $H\times W\times C$ to a feature map of size $\frac{H}{2}\times \frac{W}{2}\times D_1$. Immediately thereafter, a Wavelet-based Mamba Block (WMB) is applied to capture global low-frequency information. In the second stage, a patch embedding layer with a $2\times2$ patch size projects the features to a resolution of $\frac{H}{4}\times \frac{W}{4}$, following the design of VMamba~\cite{zhu2024vision}. Subsequent stages incorporate a patch merging layer for additional $2\times$ down-sampling and several SCVSS modules for advanced feature extraction. The feature dimensions across the five stages are set as $D=\{48,\,96,\,192,\,384,\,768\}$, and the numbers of SCVSS modules per stage (from stage 2 to stage 5) are $\{2,\,2,\,5,\,2\}$, respectively. Pre-trained weights from VMambaV2 are used to initialize the SnakeVSS blocks and patch merging layers, while the patch embedding block is trained from scratch due to differences in patch size and input channels.

\subsubsection{SCVSS Module}

At the heart of the encoder lies the SCVSS module, which integrates three parallel branches: a conventional convolution branch to capture local features, a VSS branch to model horizontal and vertical relationships and a SnakeVSS branch that reorders feature patches along serpentine directions to capture curvilinear structures more effectively. Details of SnakeVSS and VSS structure are depicted in Fig.~\ref{fig:fig4}. This combination allows the module to capture both fine-grained and long-range dependencies, which are crucial for accurate segmentation of complex anatomical structures. For an input feature map $x$, the three branches are computed as:
\begin{equation}
x_{\text{conv}} = \operatorname{CONV}\bigl(\operatorname{Norm}(x)\bigr),
\end{equation}
\begin{equation}
x_{\text{snakevss}} = \operatorname{SnakeVSS}(x),
\end{equation}
\begin{equation}
x_{\text{vss}} = \operatorname{VSS}(x).
\end{equation}

Each branch output is refined through a Spatial and Channel Attention (SCA) mechanism and then aggregated with the input using a residual connection. An MLP with DropPath regularization further processes the combined features:
\begin{equation}
\resizebox{0.67\hsize}{!}{$
x_{\text{out}} = \operatorname{MLP}\Bigl(x + \operatorname{DropPath}\Bigl(\operatorname{SCA}(x_{\text{conv}}) + \operatorname{SCA}(x_{\text{snakevss}}) + \operatorname{SCA}(x_{\text{vss}})\Bigr)\Bigr).
$}
\end{equation}

The SnakeVSS branch refines the scanning process by reordering feature patches according to serpentine patterns. Unlike conventional scanning directions ($v\in\{1,2,3,4\}$), the SnakeVSS branch defines new serpentine directions ($v_s\in\{s1, s2, s3, s4\}$), which allows better capture of curvilinear structures. This reordering process is formulated as follows:
\begin{align}
x_{\nu},\, x_{\nu_s} &= \operatorname{expand}(x,\, v,\, v_s),\\[1mm]
\overline{x_v},\, \overline{x_{v_s}} &= \operatorname{S6}(x_v,\, x_{v_s}),\\[1mm]
\overline{x_{\nu}} &= \operatorname{merge}\Bigl(\overline{x_1},\, \overline{x_2},\, \overline{x_3},\, \overline{x_4}\Bigr),\\[1mm]
\overline{x_{\nu_s}} &= \operatorname{merge}\Bigl(\overline{x_{s1}},\, \overline{x_{s2}},\, \overline{x_{s3}},\, \overline{x_{s4}}\Bigr).
\end{align}
The \texttt{expand} and \texttt{merge} operations split and recombine the feature map into sequences, while the S6 module forms the core of the Mamba operation, allowing each element to interact with previously scanned elements.

The SCA mechanism further refines the aggregated features by combining spatial and channel attention, as depicted in Fig.~\ref{fig:fig5}. 
Spatial attention is computed by applying both max pooling and average pooling, followed by fully connected layers:
\begin{align}
x_{s_{\max}} &= \mathrm{FC}\bigl(\mathrm{ReLU}(\mathrm{FC}(\mathrm{MaxPooling}(x)))\bigr),\\[1mm]
x_{s_{\text{avg}}} &= \mathrm{FC}\bigl(\mathrm{ReLU}(\mathrm{FC}(\mathrm{AveragePooling}(x)))\bigr),\\[1mm]
x_{s_{\text{output}}} &= x \odot \mathrm{Sigmoid}\bigl(x_{s_{\max}} + x_{s_{\text{avg}}}\bigr).
\end{align}
Channel attention is then achieved by concatenating the channel-wise maximum and average of $x_{s_{\text{output}}}$, processing the result with a convolution, and applying a sigmoid activation:
\begin{equation}
\resizebox{0.66\hsize}{!}{$
\begin{aligned}
x_{c} &= \mathrm{Conv}\Bigl(\mathrm{Concat}\bigl(\mathrm{MaxChannel}(x_{s_{\text{output}}}),\, \mathrm{MeanChannel}(x_{s_{\text{output}}})\bigr)\Bigr).
\end{aligned}
$}
\end{equation}

\begin{equation}
\resizebox{0.37\hsize}{!}{$
\begin{aligned}
x_{\text{output}} &= x_{s_{\text{output}}} \odot \mathrm{Sigmoid}\bigl(x_{c}\bigr).
\end{aligned}
$}
\end{equation}

\subsubsection{Wavelet-based Mamba Block (WMB)}

\begin{figure}[htbp]
\setlength{\abovecaptionskip}{2pt}
\setlength{\belowcaptionskip}{0pt}
\centering
\includegraphics[width=0.5\textwidth]{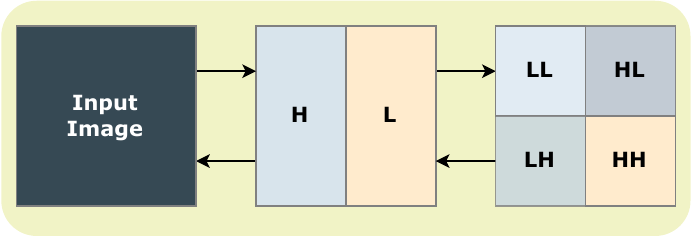}
\caption{Schematic diagram of wavelet decomposition.}
\label{fig:fig3}
\end{figure}

To further enhance global context, selected encoder stages integrate the Wavelet-based Mamba Block. Given an input feature map $x\in\mathbb{R}^{H\times W\times C}$, WMB first applies LayerNorm and then performs a 2D discrete wavelet transform to decompose $x$ into a low-frequency component $F_{LL}$ and three high-frequency components $\{F_{LH},\,F_{HL},\,F_{HH}\}$, as depicted in Fig.~\ref{fig:fig3}. The low-frequency branch processes $F_{LL}$ with a $3\times3$ convolution and employs a Channel-wise Mamba module to capture long-range dependencies, while the high-frequency sub-bands are processed by shallow convolutions. An inverse wavelet transform (IWT) reconstructs the refined features. Formally, the operations are:
\begin{align}
I' &= \operatorname{WM}\bigl(\operatorname{LN}(x)\bigr) + x,\\[1mm]
I'' &= \operatorname{FFN}\bigl(\operatorname{LN}(I')\bigr) + I',
\end{align}
where WM denotes the wavelet-based Mamba operation and FFN is a feed-forward network. 

\subsection{Decoder}

The decoder recovers the spatial resolution and fuses multi-scale information using up-sampling blocks. Initially, feature maps extracted from the encoder are adjusted via simple convolutional blocks to align their channel dimensions. These features are then concatenated with outputs from the preceding decoder stage through skip connections and further fused using additional convolutional layers. Deep supervision is imposed at multiple scales by employing auxiliary segmentation heads (implemented as $1\times1$ convolutions) to generate intermediate segmentation outputs. The final segmentation head comprises a two-layer convolutional block that fuses the multi-scale features, followed by a $1\times1$ convolution to produce the final segmentation map.

\subsection{Loss Function and Training}

The network is trained end-to-end using a composite loss function that combines Dice loss and Cross-Entropy loss to address both region overlap and pixel-wise classification accuracy:
\begin{equation}
\mathcal{L}_{\text{seg}} = \mathcal{L}_{\text{Dice}} + \mathcal{L}_{\text{CE}},
\end{equation}
where:
\begin{itemize}
    \item $\mathcal{L}_{\text{Dice}}$ measures the overlap between the predicted segmentation $\widehat{Y}$ and the ground truth $Y$:
    \begin{equation}
    \mathcal{L}_{\text{Dice}} = 1 - \frac{2 \sum_{i} \widehat{Y}_i \cdot Y_i}{\sum_{i} \widehat{Y}_i + \sum_{i} Y_i},
    \end{equation}
    \item $\mathcal{L}_{\text{CE}}$ computes the pixel-wise classification error using Cross-Entropy:
    \begin{equation}
    \mathcal{L}_{\text{CE}} = -\sum_{i} Y_i \cdot \log(\widehat{Y}_i).
    \end{equation}
\end{itemize}

Pre-trained weights from VMambaV2 are used to initialize the SnakeVSS blocks and patch merging layers, while the remaining modules are trained from scratch. Optimization is performed using AdamW with a cosine annealing learning rate scheduler, which dynamically adjusts the learning rate to improve convergence and avoid local minima.

In summary, our method effectively combines Mamba-inspired modules, a novel snake scanning strategy, and wavelet-based operations within a streamlined encoder-decoder architecture, achieving state-of-the-art segmentation performance with significantly reduced computational complexity.

\section{Experiments}

\subsection{Datasets}

\subsubsection{Private Datasets}
Our dataset consists of anonymized CT scans of patients with recurrent NPC at the Eye \& ENT Hospital of Fudan University as part of routine clinical imaging examinations. These scans, with a slice thickness of approximately 5.0\, mm, cover the nasopharynx and adjacent anatomical structures that are critical for treatment planning in the postoperative setting. Unlike primary tumor segmentation, our focus is on the precise delineation of several key airway-related structures that must be carefully spared during subsequent airway risk assessment. 
All images were manually annotated by experienced radiation oncologists and radiologists, ensuring consensus-driven segmentation of critical regions, including the infraglottic larynx, supraglottic larynx, pharyngeal constrictors, oropharynx, nasopharynx, and larynx-glottic. These regions were selected due to their essential roles in maintaining vocal cord function, swallowing, and speech, and because their accurate segmentation is vital for evaluating potential airway risks in postoperative scenarios. Prior to model training, the CT images underwent standardized preprocessing—including noise removal, intensity normalization, and resizing—to ensure consistency across the dataset and enhance model performance. 
The corresponding label maps, serve as the ground truth for supervised training.
This study was approved by the Ethics Committee of the Eye \& ENT Hospital of Fudan University (Approval No. 2024232).

\subsubsection{SegRap 2023 Challenge Public Dataset}
The SegRap 2023 Challenge public dataset consists of CT scans collected from 120 NPC patients prior to treatment. These scans were acquired using Siemens CT scanners with a tube voltage of 120\,kV, a tube current of 300\, mA, a slice thickness of 3.0\, mm, and resolutions of either 1024×1024 or 512×512 pixels. The dataset includes both contrast-enhanced and non-contrast-enhanced head and neck CT scans; the contrast-enhanced images were obtained using iohexol (administered at 60–80\, mL with an injection rate of 2\,mL/s, without any delay), thereby providing detailed anatomical information critical for NPC assessment and treatment planning.
Although each CT image in the dataset is accompanied by manual segmentations of 45 structures and two gross tumor volumes (GTVs), for this study we exclusively focus on the segmentation of the trachea. The trachea is a vital structure responsible for maintaining respiratory function and plays a critical role in postoperative airway risk assessment. By focusing on the trachea, our work aims to evaluate the performance of our TopoWMamba model in accurately segmenting this critical anatomical region, thereby providing a foundation for improved postoperative airway risk assessment and management in NPC patients.

\subsection{Experimental Setup}

\subsubsection{Implementation Details}

We implement TopoWMamba in PyTorch, using efficient 2D convolutions and wavelet transformations. The Haar wavelet decompositions are computed using off-the-shelf differentiable wavelet transform layers. We train the model using the Adam optimizer with a learning rate of $1 \times 10^{-4}$, decaying it slowly as training progresses. Typical training involves $100$ epochs, with early stopping based on validation performance. We train our model on NVIDIA GeForce RTX 3090 with 24 GB memory. During the training period, the batch size is set as 4.

\subsubsection{Baselines and State-of-the-Art Comparisons}
To evaluate the performance of TopoWMamba, we compare it against several SOTA networks and baseline models. These baselines cover a range of architectures, from traditional encoder-decoder models to more advanced networks incorporating attention mechanisms and transformers, as well as recent developments utilizing novel backbones and architectures. The selected baselines are as follows:

\begin{itemize} \item \textbf{Attention UNet} \cite{oktay2018attention}: This model integrates attention mechanisms to focus on salient regions of the image, which helps to improve segmentation accuracy in areas with complex structures.
\item \textbf{FPN with ResNet} \cite{seferbekov2018feature}: The Feature Pyramid Network (FPN) with ResNet backbone leverages multi-scale feature extraction through lateral connections, making it effective for capturing fine-grained details in medical image segmentation tasks.

\item \textbf{UNet++} \cite{zhou2018unet++}: An extension of U-Net, U-Net++ introduces dense skip pathways, improving the flow of feature maps between encoder and decoder, which enhances the model’s ability to recover fine segmentation details.

\item \textbf{SegNet} \cite{badrinarayanan2017segnet}: A deep convolutional encoder-decoder architecture, SegNet features efficient upsampling layers that allow for pixel-wise segmentation with minimal computational cost, making it a strong baseline for comparison.

\item \textbf{TransUNet} \cite{chen2021transunet}: This model combines convolutional neural networks (CNNs) with transformers, enabling it to capture both local and global contextual information, thus improving segmentation performance in tasks requiring long-range dependencies.

\item \textbf{SwinUNet} \cite{cao2022swin}: Utilizing the Swin Transformer as a backbone, SwinUNet combines local patch-based attention with hierarchical feature extraction, making it highly effective for medical image segmentation tasks where spatial context and fine details are crucial.

\item \textbf{MambaUNet} \cite{wang2024mamba}: The model adopts VMamba-based structure, infused with skip connections to preserve spatial information across different scales of the network. This design facilitates a comprehensive feature learning process, capturing intricate details and broader semantic contexts within medical images.

\item \textbf{UNet} \cite{ronneberger2015u}: A widely-used encoder-decoder architecture that serves as a solid baseline for many segmentation tasks, providing a simple yet effective framework for medical image segmentation.

\item \textbf{R2U-Net} \cite{alom2018recurrent}: This model enhances the traditional U-Net by incorporating recurrent residual connections. It effectively captures multi-scale contextual information and improves feature propagation, making it particularly well-suited for complex medical imaging tasks.
\end{itemize}

These baseline models provide a comprehensive benchmark for assessing the effectiveness of the novel components in TopoWMamba, such as Wavelet-Mamba Blocks (WMB) and the integration of advanced feature extraction techniques. By comparing TopoWMamba against these models, we aim to demonstrate the improvements in segmentation performance brought by the unique design of TopoWMamba, especially in terms of multi-scale feature extraction and boundary refinement.

\subsubsection{Evaluation Metrics}

We employ multiple metrics to evaluate segmentation performance comprehensively:
\begin{itemize}
    \item \textbf{Dice Similarity Coefficient (Dice(\%))}: Measures the overlap between predicted and ground truth:
    \begin{equation}
        \text{Dice}=\frac{2|A \cap B|}{|A|+|B|},
    \end{equation}
    where $A$ and $B$ represent the predicted and ground truth segmentation regions, respectively.
    \item \textbf{Hausdorff Distance at 95\% (HD95(mm))}: Measures the maximum distance between predicted and ground truth boundaries, considering the 95th percentile of distances to reduce sensitivity to outliers:
    \begin{equation}
       \text{HD95}=\max \left\{\sup _{a \in A} \min _{b \in B} d(a, b), \sup _{b \in B} \min _{a \in A} d(a, b)\right\},
    \end{equation}
    where $d(a, b)$ is the Euclidean distance between points $a$ and $b$.
    \item \textbf{mean Intersection over Union (mIoU(\%))}: Calculates the average intersection over union for each class, providing a measure of the overall segmentation accuracy:
    \begin{equation}
        \text{mIoU}=\frac{1}{C} \sum_{c=1}^C \frac{\left|A_c \cap B_c\right|}{\left|A_c \cup B_c\right|},
    \end{equation}
    where $C$ represents the number of classes, and $A_c$ and $\boldsymbol{B}_c$ are the predicted and ground truth regions for class $c$.
\end{itemize}

% Table generated by Excel2LaTeX from sheet 'table1 Private'
\begin{table*}[htbp]
  \centering
  \caption {Segmentation performance comparison for NPCSegCT dataset across different methods.}
  \resizebox{\linewidth}{!}{
    \begin{tabular}{cccccccccccccccccccccc}
    \toprule
    \multirow{2}[4]{*}{\textbf{Method Description}} & \multicolumn{3}{c}{\textbf{larynx-glottic}} & \multicolumn{3}{c}{\textbf{oropharynx}} & \multicolumn{3}{c}{\textbf{nasopharynx}} & \multicolumn{3}{c}{\textbf{infraglottic larynx}} & \multicolumn{3}{c}{\textbf{pharynxConst}} & \multicolumn{3}{c}{\textbf{supraglottic larynx}} & \multicolumn{3}{c}{\textbf{Mean}} \\
\cmidrule(r){2-4} \cmidrule(r){5-7} \cmidrule(r){8-10} \cmidrule(r){11-13} \cmidrule(r){14-16} \cmidrule(r){17-19} \cmidrule(r){20-22}         & \textbf{Dice} & \textbf{HD95} & \textbf{mIoU} & \textbf{Dice} & \textbf{HD95} & \textbf{mIoU} & \textbf{Dice} & \textbf{HD95} & \textbf{mIoU} & \textbf{Dice} & \textbf{HD95} & \textbf{mIoU} & \textbf{Dice} & \textbf{HD95} & \textbf{mIoU} & \textbf{Dice} & \textbf{HD95} & \textbf{mIoU} & \textbf{Dice} & \textbf{HD95} & \textbf{mIoU} \\
    \midrule
    SwinUNet & 79.48  & 10.14  & 71.98  & 79.29  & 5.50  & 75.16  & 72.58  & 2.86  & 67.86  & 82.42  & 4.41  & 75.19  & 90.30  & 3.87  & 85.31  & 85.95  & 1.73  & 83.04  & 81.67  & 4.75  & 76.42  \\
    TransUNet & 88.16  & 7.70  & 81.31  & 77.55  & 6.34  & 73.48  & 70.49  & 1.52  & 66.77  & 87.24  & 4.55  & 79.49  & 91.72  & 3.93  & 86.72  & 90.70  & 1.69  & 87.80  & 84.31  & 4.29  & 79.26  \\
    UNet & 89.01  & 7.52  & 81.61  & 82.05  & 6.34  & 76.67  & 79.85  & 4.35  & 73.81  & 86.38  & 5.63  & 77.98  & 85.83  & 5.69  & 79.59  & 86.73  & 1.84  & 83.79  & 84.98  & 5.23  & 78.91  \\
    SegNet & 85.34  & 9.01  & 77.51  & 78.61  & 6.26  & 73.61  & 53.61  & 3.34  & 50.26  & 83.01  & 6.11  & 73.61  & 86.88  & 4.96  & 81.08  & 88.50  & 1.27  & 85.71  & 79.33  & 5.16  & 73.63  \\
    Attention UNet & 88.30  & 8.80  & 74.81  & 88.12  & 9.45  & 75.39  & 50.35  & 10.57  & 42.84  & 86.15  & 6.23  & 77.92  & 88.97  & 4.85  & 71.27  & 71.27  & 5.79  & 65.70  & 78.86  & 7.61  & 67.99  \\
    UNet++ & 86.99  & 7.77  & 80.25  & 82.54  & 5.76  & 78.74  & 61.18  & 6.31  & 57.17  & 87.91  & 4.72  & 79.93  & 90.96  & 3.93  & 86.04  & 89.45  & 1.98  & 86.86  & 83.17  & 5.08  & 78.17  \\
    MambaUNet & 63.41  & 11.16  & 54.75  & 72.98  & 10.35  & 64.45  & 81.27  & 3.66  & 74.59  & 83.45  & 6.67  & 74.29  & 82.90  & 9.23  & 74.84  & 81.26  & 3.98  & 77.22  & 77.55  & 7.51  & 70.02  \\
    FPN+ResNet & 89.62  & 6.22  & 83.37  & 78.78  & 6.45  & 74.20  & 73.06  & 3.13  & 68.70  & 85.49  & 4.21  & 77.31  & 90.58  & 3.72  & 85.19  & 88.71  & 1.60  & 85.76  & 84.37  & 4.22  & 79.09  \\
    R2U-Net & 86.33  & 7.09  & 79.53  & 85.54  & 7.10  & 81.16  & 65.62  & 2.19  & 62.97  & 87.60  & 2.81  & 80.89  & 91.42  & 3.66  & 86.77  & 89.74  & 2.49  & 86.86  & 84.38  & 4.22  & 79.70  \\
    \rowcolor{lightblue} TopoWMamba & 91.11  & 6.15  & 84.84  & 83.71  & 5.25  & 79.58  & 82.32  & 4.21  & 77.85  & 90.66  & 3.15  & 83.88  & 90.85  & 3.49  & 86.33  & 89.47  & 2.43  & 87.00  & 88.02  & 4.11  & 83.25  \\
    \bottomrule
    \end{tabular}%
    }
  \label{tab:table1}%
  \vspace{-.1in}
\end{table*}%

\begin{figure*}[htbp]
\centering
\includegraphics[width=6.6in]{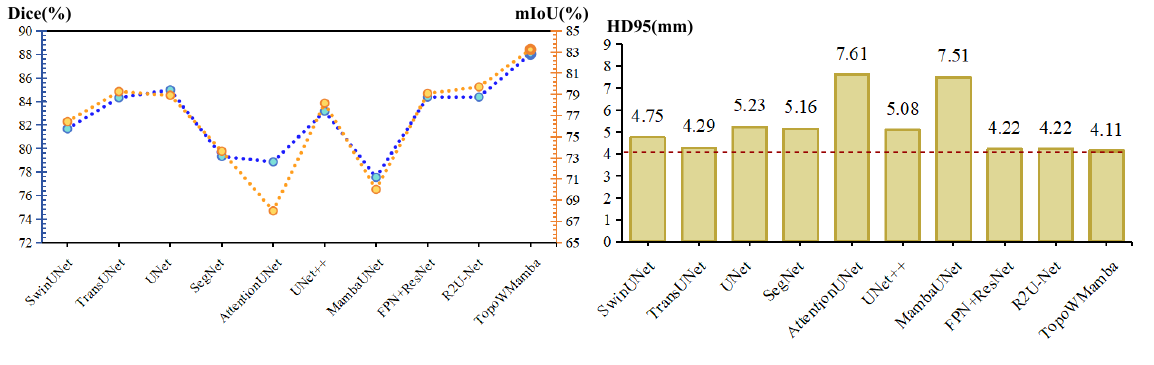}
\caption{The segmentation performance comparison for NPCSegCT dataset across different methods, including average Dice, average mIoU, and average HD95 on 6 regions of interest.}
\label{fig:fig6}
\end{figure*}

\subsection{Quantitative Results}

\subsubsection{Overall Performance}
\begin{figure*}[htbp]
\setlength{\abovecaptionskip}{0pt}
\setlength{\belowcaptionskip}{0pt}
\centering
\includegraphics[width=6.6in]{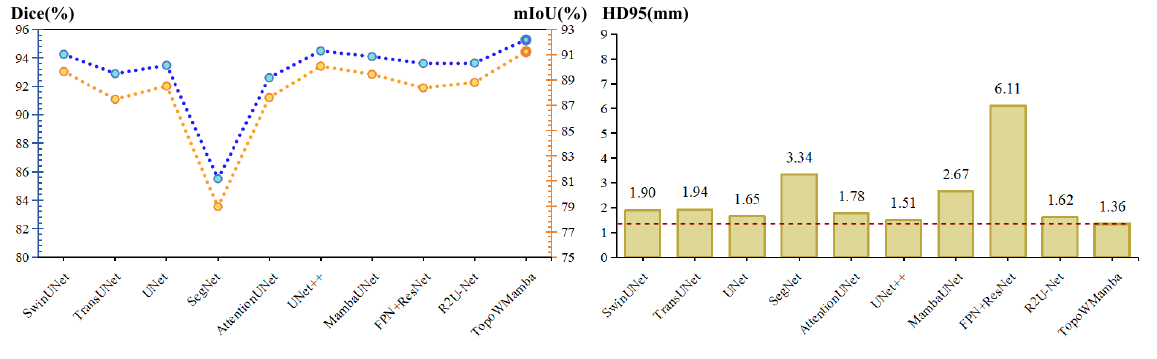}
\caption{The segmentation performance comparison for SegRap 2023 challenge public dataset across different methods, including average Dice, average mIoU, and average HD95 on Trachea.}
\label{fig:fig7}
\end{figure*}

The results, summarized in Table~\ref{tab:table1}, Table~\ref{tab:table2} and Fig.~\ref{fig:fig6}, Fig.~\ref{fig:fig7}, demonstrate that TopoWMamba consistently outperforms all baseline models in terms of key evaluation metrics on NPCSegCT dataset and SegRap 2023 challenge public dataset.

% Table generated by Excel2LaTeX from sheet 'table2 Segrap'
\begin{table*}[htbp]
  \centering
  \small
  \caption{Segmentation performance comparison for SegRap 2023 challenge public dataset across different methods.}
  \scalebox{1.00}{
    \begin{tabular}{cccc}
    \toprule
    \multirow{2}[4]{*}{\textbf{Method Description}} & \multicolumn{3}{c}{\textbf{Trachea}} \\
\cmidrule{2-4}         & \multicolumn{1}{c}{\textbf{Dice}} & \multicolumn{1}{c}{\textbf{HD95}} & \multicolumn{1}{c}{\textbf{mIoU}} \\
    \midrule
    SwinUNet & 94.25  & 1.90  & 89.66  \\
    TransUNet & 92.89  & 1.94  & 87.48  \\
    UNet & 93.47  & 1.65  & 88.52  \\
    SegNet & 85.50  & 3.34  & 78.99  \\
    AttentionUNet & 92.60  & 1.78  & 87.61  \\
    UNet++ & 94.49  & 1.51  & 90.09  \\
    MambaUNet & 94.09  & 2.67  & 89.45  \\
    FPN+ResNet & 93.60  & 6.11  & 88.38  \\
    R2U-Net & 93.63  & 1.62  & 88.81  \\
    \rowcolor{lightblue} TopoWMamba & 95.26  & 1.36  & 91.24  \\
    \bottomrule
    \end{tabular}%
  }
  \label{tab:table2}%
  \vspace{-.1in}
\end{table*}%

On the NPCSegCT dataset, TopoWMamba achieves the highest mean Dice score of 88.02\%, outperforming the second-best method, FPN with ResNet, by a notable margin. Furthermore, TopoWMamba achieves superior HD95 and mIoU scores across all regions of interest (ROIs), including the larynx-glottic, oropharynx, nasopharynx, and others. For example, in the larynx-glottic region, TopoWMamba achieves a Dice score of 91.11\%, significantly higher than UNet and other competing methods. In terms of HD95, TopoWMamba delivers a sharp reduction in boundary errors (4.11 mm), indicating improved precision in delineating organ boundaries, which is crucial for assessing postoperative airway risk. These results demonstrate TopoWMamba’s capability in accurately segmenting the critical airway-related structures.

On the SegRap 2023 challenge public dataset, specifically for trachea segmentation, TopoWMamba again shows remarkable performance, achieving the highest Dice score of 95.26\% and the lowest HD95 of 1.36 mm. This outperforms other methods, including UNet and TransUNet, by a clear margin. The improved segmentation performance is particularly evident in the high precision of boundary delineations, as indicated by the lower HD95 values. TopoWMamba's effective use of multi-branch high-frequency extraction and topology-aware design contributes to its superior performance in these segmentation tasks.

Overall, the experimental results highlight the advantages of TopoWMamba in airway-related structure segmentation across different datasets. TopoWMamba’s integration of wavelet decomposition, frequency-domain analysis, and topology-informed architecture allows it to achieve more accurate and stable segmentations, particularly in challenging regions with complex anatomical structures. These precise margin identification and accurate delineation of critical structures are crucial for accurate postoperative airway risk assessment.

\begin{figure*}[htbp]
\setlength{\abovecaptionskip}{2pt}
\setlength{\belowcaptionskip}{0pt}
\centering
\includegraphics[width=\textwidth]{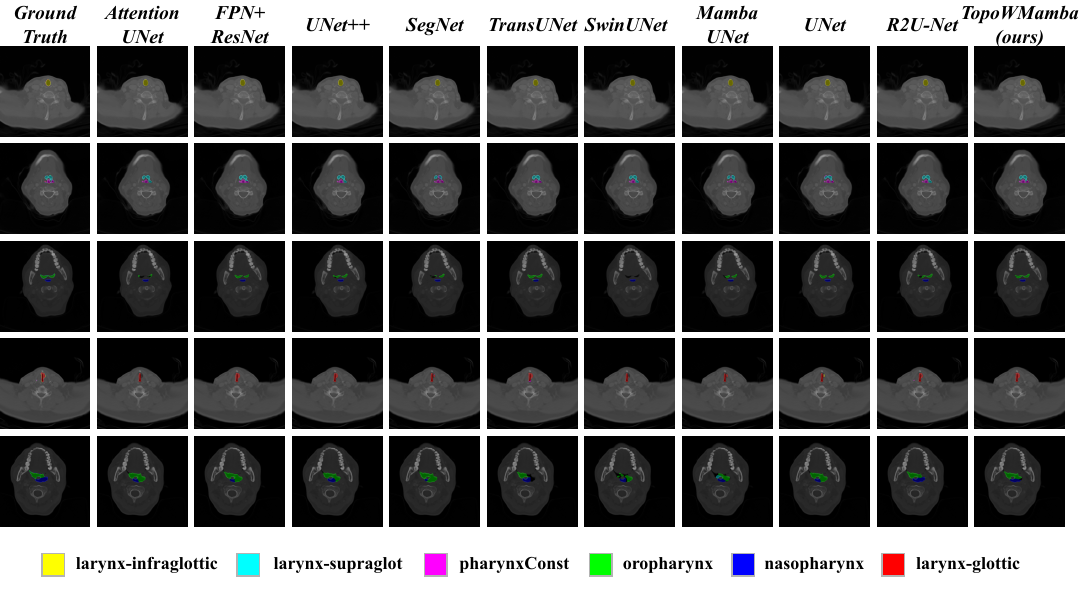}
\caption{Qualitative comparison of segmentation results for TopoWMamba and other methods on NPCSegCT dataset. Columns represent various models. Each row corresponds to a region of interest.}
\label{fig:fig8}
\end{figure*}

\subsubsection{Comparisons with Baselines.}
Compared to the baseline method, UNet, TopoWMamba demonstrates a significant improvement in segmentation performance. By incorporating wavelet decomposition and selective state space modeling, TopoWMamba achieves a substantial boost in Dice and mIoU and a noticeable reduction in boundary errors. This is particularly beneficial in delineating complex anatomical structures with precision, which is crucial for treatment planning. Furthermore, TopoWMamba’s unique multi-branch high-frequency extraction and topology-aware design allow for better fine-grained segmentation, further improving the segmentation quality, especially in regions such as the larynx, oropharynx, and nasopharynx.

When compared to Attention UNet, TopoWMamba’s frequency-domain analysis and channel attention mechanisms provide more targeted enhancement of relevant features. While Attention UNet uses general attention mechanisms, TopoWMamba’s focused frequency-domain analysis and selective attention mechanisms effectively extract high-frequency components, improving both segmentation accuracy and stability, particularly in the critical ROIs.

Although methods like R2U-Net and TransUNet show strong baseline performances, TopoWMamba refines segmentation results even further. The multi-branch structure of TopoWMamba, combined with its frequency-guided and topology-aware architecture, surpasses these methods, demonstrating a clear advantage in fine-tuning segmentation. The topology-aware design ensures that the network maintains consistent topological relationships, further enhancing the accuracy of boundary delineation.

\begin{figure*}[htbp]
\setlength{\abovecaptionskip}{2pt}
\setlength{\belowcaptionskip}{0pt}
\centering
\includegraphics[width=\textwidth]{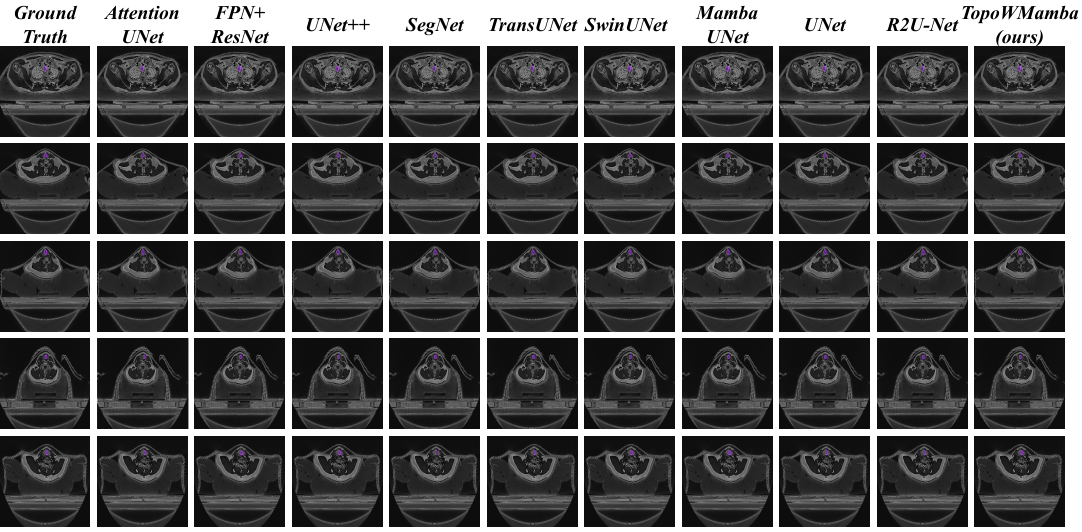}
\caption{Qualitative comparison of segmentation results for TopoWMamba and other methods on SegRap 2023 challenge public dataset. Columns represent various models. Each row corresponds to a CT slice.}
\label{fig:fig9}
\end{figure*}

\subsubsection{Qualitative Analysis}

Fig.~\ref{fig:fig8} and Fig.~\ref{fig:fig9} presents visual examples of segmentation results from TopoWMamba and other methods. TopoWMamba’s predictions closely align with the expert-annotated ground truth, especially around complex anatomical structures and narrow regions. For example, TopoWMamba accurately delineates critical structures such as the larynx, oropharynx, and nasopharynx, demonstrating its ability to integrate both global and fine-grained details.

Moreover, TopoWMamba excels in handling difficult and ambiguous regions. By integrating low-level and high-level information through frequency-domain and topology-aware guidance, TopoWMamba generates robust segmentation results even in the presence of complex, overlapping structures, maintaining anatomical consistency. This makes TopoWMamba particularly well-suited for the segmentation of airway-related structures.

\subsection{Ablation Studies}

In this section, we present a series of ablation experiments conducted to assess the contributions of different components in TopoWMamba. Specifically, we investigate the effects of WMB in the encoder and decoder, the placement of WMB in the encoder, and the impact of the SnakeVSS block. These experiments provide insights into how each part of the architecture contributes to the overall segmentation performance.

\subsubsection{Effect of WMB in Encoder and Decoder}

\begin{figure}[htbp]
\setlength{\abovecaptionskip}{2pt}
\setlength{\belowcaptionskip}{0pt}
\centering
\includegraphics[width=0.5\textwidth]{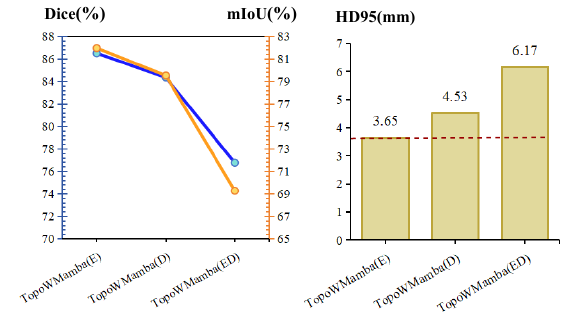}
\caption{The segmentation performance comparison for NPCSegCT dataset across different experiment configurations, including average Dice, average mIoU, and average HD95 on 6 regions of interest.}
\label{fig:fig10}
\end{figure}

Firstly, we explore the effect of placing WMB blocks at different stages of the encoder and decoder. The results of these experiments are presented in Table \ref{tab:table3} and Fig.~\ref{fig:fig10}. We observe that the configuration where WMB is applied in both the encoder and decoder performs the worst, achieving a Dice score of 76.78\%, an HD95 of 6.17 mm, and a mIoU of 69.29\%. This is in contrast to the first experiment, where adding WMB only in the encoder results in significantly better performance, with a Dice score of 86.51\%, HD95 of 3.65 mm, and mIoU of 81.97\%. Adding WMB exclusively to the decoder also yields strong results, with a Dice score of 84.35\%, HD95 of 4.53 mm, and mIoU of 79.53\%. These findings suggest that applying WMB in the encoder helps improve multi-scale feature extraction while adding WMB in the decoder helps refine boundary details, but combining them might introduce complexity that negatively impacts the performance.

\begin{table*}[htbp]
  \centering
  \small
  \caption{Ablation study of WMB placement strategies. TopoWMamba(E): WMB in Encoder only; TopoWMamba(D): WMB in Decoder only; TopoWMamba(ED): WMB in both Encoder \& Decoder.}
  % \ding{172}WMB only added after each stage in the encoder. \ding{173}WMB only added after each stage in the encoder. \ding{174}WMB added after each stage in both the encoder and decoder.}
  \resizebox{\linewidth}{!}{
    \begin{tabular}{cccccccccccccccccccccc}
    \toprule
    \multirow{2}[4]{*}{\textbf{Model Variant}} & \multicolumn{3}{c}{\textbf{larynx-glottic}} & \multicolumn{3}{c}{\textbf{oropharynx}} & \multicolumn{3}{c}{\textbf{nasopharynx}} & \multicolumn{3}{c}{\textbf{infraglottic larynx}} & \multicolumn{3}{c}{\textbf{pharynxConst}} & \multicolumn{3}{c}{\textbf{supraglottic larynx}} & \multicolumn{3}{c}{\textbf{Mean}} \\
\cmidrule(r){2-4} \cmidrule(r){5-7} \cmidrule(r){8-10} \cmidrule(r){11-13} \cmidrule(r){14-16} \cmidrule(r){17-19} \cmidrule(r){20-22} 
& \textbf{Dice} & \textbf{HD95} & \textbf{mIoU} & \textbf{Dice} & \textbf{HD95} & \textbf{mIoU} & \textbf{Dice} & \textbf{HD95} & \textbf{mIoU} & \textbf{Dice} & \textbf{HD95} & \textbf{mIoU} & \textbf{Dice} & \textbf{HD95} & \textbf{mIoU} & \textbf{Dice} & \textbf{HD95} & \textbf{mIoU} & \textbf{Dice} & \textbf{HD95} & \textbf{mIoU} \\
    \midrule
    TopoWMamba(E)    & 90.74  & 6.27  & 84.63  & 83.81  & 5.06  & 79.94  & 77.24  & 1.80  & 73.91  & 87.94  & 3.57  & 81.06  & 91.20  & 3.74  & 86.43  & 88.15  & 1.48  & 85.86  & 86.51  & 3.65  & 81.97  \\
    TopoWMamba(D)    & 89.76  & 8.88  & 83.29  & 79.93  & 6.68  & 75.84  & 69.03  & 1.96  & 65.40  & 87.02  & 4.82  & 79.53  & 91.48  & 3.53  & 86.68  & 88.88  & 1.31  & 86.42  & 84.35  & 4.53  & 79.53  \\
    TopoWMamba(ED)   & 87.10  & 6.78  & 80.06  & 80.17  & 6.07  & 75.37  & 68.86  & 5.02  & 59.42  & 65.06  & 9.93  & 54.77  & 75.79  & 6.58  & 68.46  & 83.67  & 2.64  & 77.64  & 76.78  & 6.17  & 69.29  \\
    \bottomrule
    \end{tabular}%
  }
  \label{tab:table3}%
\end{table*}%

\subsubsection{Impact of WMB Placement in the Encoder}

We investigate the impact of placing WMB at different stages in the encoder, with the decoder configuration kept constant in two distinct experimental setups. The first set of experiments involves adding WMB after each stage in the decoder, while the second set involves no WMB in the decoder, focusing only on the encoder configuration.

The results of the first set of experiments, shown in Table \ref{tab:table4}, indicate that placing WMB after the 2nd and 4th stages of the encoder yields the best performance, with a Dice score of 85.90\%, HD95 of 3.98 mm, and mIoU of 80.48\%. This configuration enhances the model's ability to capture multi-scale features across multiple stages of the encoder. Meanwhile, adding WMB  after the 1st, 3rd, and 5th stages results in slightly lower performance, with a Dice score of 83.34\%, HD95 of 4.15 m,m and mIoU of 77.70\%, highlighting that the distribution of WMB across multiple encoder stages is beneficial for segmentation accuracy.

\begin{figure*}[htbp]
\setlength{\abovecaptionskip}{2pt}
\setlength{\belowcaptionskip}{0pt}
\centering
\includegraphics[width=15cm]{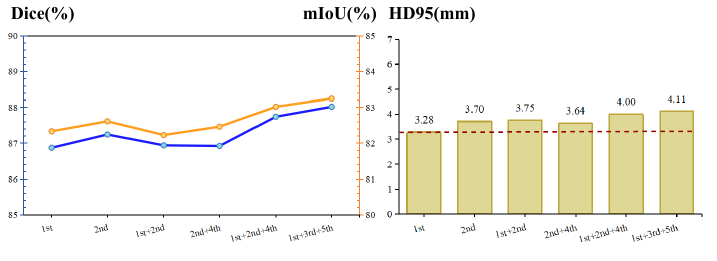}
\caption{The segmentation performance comparison for NPCSegCT dataset across different experiment configurations, including average Dice, average mIoU, and average HD95 on 6 regions of interest.}
\label{fig:fig11}
\end{figure*}

In the second set of experiments, where WMB is not added in the decoder (Table \ref{tab:table5} and Fig.~\ref{fig:fig11}), placing WMB after specific stages in the encoder reveals a similar trend. The configuration where WMB is applied after the 1st, 3rd, and 5th encoder stages still performs well, with a Dice score of 88.02\%, HD95 of 4.11 mm, and mIoU of 83.25\%. However, adding WMB only after the 2nd stage or the 1st and 2nd stages results in slightly lower performance, indicating that limiting WMB to fewer stages reduces the model’s ability to effectively capture multi-scale information.

These results confirm that strategically placing WMB at specific stages in the encoder enhances feature extraction and segmentation accuracy.

\begin{table*}[htbp]
  \centering
  \small
  \caption{Ablation study results for evaluating the impact of WMB placement in the encoder when WMB is placed after each layer in the decoder.}
  \resizebox{\linewidth}{!}{
    \begin{tabular}{cccccccccccccccccccccc}
    \toprule
    \multirow{2}[4]{*}{\textbf{Encoder layer with WMB}} & \multicolumn{3}{c}{\textbf{larynx-glottic}} & \multicolumn{3}{c}{\textbf{oropharynx}} & \multicolumn{3}{c}{\textbf{nasopharynx}} & \multicolumn{3}{c}{\textbf{infraglottic larynx}} & \multicolumn{3}{c}{\textbf{pharynxConst}} & \multicolumn{3}{c}{\textbf{supraglottic larynx}} & \multicolumn{3}{c}{\textbf{Mean}} \\
\cmidrule(r){2-4} \cmidrule(r){5-7} \cmidrule(r){8-10} \cmidrule(r){11-13} \cmidrule(r){14-16} \cmidrule(r){17-19} \cmidrule(r){20-22}         & \textbf{Dice(\%)} & \textbf{HD95} & \textbf{mIoU} & \textbf{Dice} & \textbf{HD95} & \textbf{mIoU} & \textbf{Dice} & \textbf{HD95} & \textbf{mIoU} & \textbf{Dice} & \textbf{HD95} & \textbf{mIoU} & \textbf{Dice} & \textbf{HD95} & \textbf{mIoU} & \textbf{Dice} & \textbf{HD95} & \textbf{mIoU} & \textbf{Dice} & \textbf{HD95} & \textbf{mIoU} \\
    \midrule
    1st+3rd+5th    & 91.78  & 6.79  & 85.54  & 79.78  & 3.54  & 76.11  & 64.46  & 5.34  & 55.56  & 85.46  & 3.95  & 78.81  & 90.37  & 3.85  & 85.11  & 88.18  & 1.41  & 85.09  & 83.34  & 4.15  & 77.70  \\
    2nd+4th    & 90.07  & 6.24  & 83.42  & 84.56  & 4.53  & 80.05  & 78.57  & 2.99  & 74.16  & 86.15  & 4.71  & 78.63  & 89.74  & 3.59  & 84.28  & 86.31  & 1.78  & 82.31  & 85.90  & 3.98  & 80.48  \\
    \bottomrule
    \end{tabular}%
  }
  \label{tab:table4}%
\end{table*}%

\begin{table*}[htbp]
  \centering
  \small
  \caption{Ablation study results for evaluating the impact of WMB placement in the encoder when there is no WMB in the decoder.}
  \resizebox{\linewidth}{!}{
    \begin{tabular}{cccccccccccccccccccccc}
    \toprule
    \multirow{2}[4]{*}{\textbf{Encoder layer with WMB}} & \multicolumn{3}{c}{\textbf{larynx-glottic}} & \multicolumn{3}{c}{\textbf{oropharynx}} & \multicolumn{3}{c}{\textbf{nasopharynx}} & \multicolumn{3}{c}{\textbf{infraglottic larynx}} & \multicolumn{3}{c}{\textbf{pharynxConst}} & \multicolumn{3}{c}{\textbf{supraglottic larynx}} & \multicolumn{3}{c}{\textbf{Mean}} \\
\cmidrule(r){2-4} \cmidrule(r){5-7} \cmidrule(r){8-10} \cmidrule(r){11-13} \cmidrule(r){14-16} \cmidrule(r){17-19} \cmidrule(r){20-22}         & \textbf{Dice} & \textbf{HD95} & \textbf{mIoU} & \textbf{Dice} & \textbf{HD95} & \textbf{mIoU} & \textbf{Dice} & \textbf{HD95} & \textbf{mIoU} & \textbf{Dice} & \textbf{HD95} & \textbf{mIoU} & \textbf{Dice} & \textbf{HD95} & \textbf{mIoU} & \textbf{Dice} & \textbf{HD95} & \textbf{mIoU} & \textbf{Dice} & \textbf{HD95} & \textbf{mIoU} \\
    \midrule
    1st    & 91.71  & 5.86  & 85.51  & 82.36  & 3.45  & 78.65  & 78.12  & 2.76  & 74.43  & 89.05  & 3.13  & 82.29  & 90.84  & 3.38  & 86.24  & 89.16  & 1.09  & 86.89  & 86.87  & 3.28  & 82.34  \\
    2nd    & 91.78  & 6.00  & 85.67  & 85.37  & 4.99  & 81.15  & 76.78  & 2.98  & 73.30  & 90.36  & 3.40  & 83.42  & 91.02  & 3.40  & 86.37  & 88.15  & 1.45  & 85.77  & 87.24  & 3.70  & 82.61  \\
    1st+2nd    & 90.37  & 5.95  & 84.28  & 86.87  & 5.58  & 82.23  & 76.86  & 2.46  & 73.48  & 88.29  & 3.71  & 81.38  & 91.60  & 3.43  & 86.90  & 87.66  & 1.40  & 85.10  & 86.94  & 3.75  & 82.23  \\
    2nd+4th    & 89.26  & 6.47  & 83.49  & 80.87  & 4.32  & 77.49  & 84.64  & 2.22  & 80.84  & 87.59  & 3.44  & 81.00  & 91.11  & 3.69  & 86.30  & 88.05  & 1.69  & 85.66  & 86.92  & 3.64  & 82.46  \\
    1st+2nd+4th    & 91.60  & 5.78  & 85.60  & 88.85  & 5.08  & 84.20  & 75.59  & 4.36  & 72.12  & 90.88  & 3.77  & 84.04  & 91.31  & 3.40  & 86.45  & 88.23  & 1.59  & 85.70  & 87.74  & 4.00  & 83.02  \\
    1st+3rd+5th    & 91.11  & 6.15  & 84.84  & 83.71  & 5.25  & 79.58  & 82.32  & 4.21  & 77.85  & 90.66  & 3.15  & 83.88  & 90.85  & 3.49  & 86.33  & 89.47  & 2.43  & 87.00  & 88.02  & 4.11  & 83.25  \\
    \bottomrule
    \end{tabular}%
  }
  \label{tab:table5}%
\end{table*}%

\subsubsection{Impact of the SnakeVSS block}
To assess the impact of the SnakeVSS block on the performance of TopoWMamba, we compared the performance of TopoWMamba with and without the SnakeVSS block. 

\begin{table*}[htbp]
  \centering
  \caption{Ablation study results for evaluating the impact of the SnakeVSS block in TopoWMamba on the NPCSegCT dataset.}
  \resizebox{\linewidth}{!}{
    \begin{tabular}{cccccccccccccccccccccc}
    \toprule
    \multirow{2}[4]{*}{\textbf{Experiment Configuration}} & \multicolumn{3}{c}{\textbf{larynx-glottic}} & \multicolumn{3}{c}{\textbf{oropharynx}} & \multicolumn{3}{c}{\textbf{nasopharynx}} & \multicolumn{3}{c}{\textbf{infraglottic larynx}} & \multicolumn{3}{c}{\textbf{pharynxConst}} & \multicolumn{3}{c}{\textbf{supraglottic larynx}} & \multicolumn{3}{c}{\textbf{Mean}} \\
\cmidrule(r){2-4} \cmidrule(r){5-7} \cmidrule(r){8-10} \cmidrule(r){11-13} \cmidrule(r){14-16} \cmidrule(r){17-19} \cmidrule(r){20-22}         & \textbf{Dice} & \textbf{HD95} & \textbf{mIoU} & \textbf{Dice} & \textbf{HD95} & \textbf{mIoU} & \textbf{Dice} & \textbf{HD95} & \textbf{mIoU} & \textbf{Dice} & \textbf{HD95} & \textbf{mIoU} & \textbf{Dice} & \textbf{HD95} & \textbf{mIoU} & \textbf{Dice} & \textbf{HD95} & \textbf{mIoU} & \textbf{Dice} & \textbf{HD95} & \textbf{mIoU} \\
    \midrule
    TopoWMamba & 91.11  & 6.15  & 84.84  & 83.71  & 5.25  & 79.58  & 82.32  & 4.21  & 77.85  & 90.66  & 3.15  & 83.88  & 90.85  & 3.49  & 86.33  & 89.47  & 2.43  & 87.00  & 88.02  & 4.11  & 83.25  \\
    TopoWMamba - SnakeVSS & 90.35 & 6.45 & 83.96 & 83.36 & 5.12 & 79.49 & 73.72 & 4.12 & 69.47 & 89.15 & 3.78 & 82.42 & 91.80 & 3.58 & 87.01 & 91.10 & 1.71 & 88.69 & 86.58 & 4.13 & 81.84 \\
    \bottomrule
    \end{tabular}%
  }
  \label{tab:table6}%
\end{table*}%

As shown in Table \ref{tab:table6}, the inclusion of the SnakeVSS block in TopoWMamba leads to notable improvements in segmentation performance. Specifically, TopoWMamba achieves higher Dice scores and mIoU values across all evaluated regions, with an overall mean Dice score of 88.02\%, compared to 86.58\% when the SnakeVSS block is removed. In addition, the HD95 metric is lower for TopoWMamba compared to the variant without SnakeVSS, indicating better boundary precision when SnakeVSS is included.

Notably, the removal of SnakeVSS leads to a decrease in performance across most anatomical regions, particularly in the larynx-glottic, oropharynx, and nasopharynx regions, where TopoWMamba without SnakeVSS shows reductions in both Dice scores and mIoU. However, for certain regions like the pharynxConst and supraglottic larynx, the performance remains relatively stable.

These results underscore the critical role of the SnakeVSS block in enhancing feature extraction and improving segmentation accuracy, particularly in challenging regions with complex anatomical structures. The performance drop observed in the absence of SnakeVSS suggests that this block significantly contributes to the model's ability to capture multi-scale features and refine segmentation boundaries, further validating its importance in our TopoWMamba architecture.

\section{Discussion and Conclusion}  

The introduction of the TopoWMamba model significantly advances the segmentation of critical airway-related structures in postoperative recurrent nasopharyngeal carcinoma (NPC) patients. Traditional segmentation techniques, such as manual contouring performed by clinicians, are both time-consuming and prone to variability. In contrast, the automated segmentation provided by TopoWMamba offers a more reliable, precise, and efficient solution. This is particularly crucial for postoperative management, where accurate segmentation of key anatomical structures, such as the airway, is necessary to predict and assess potential airway risks, such as stenosis or obstruction, which can severely affect patients' respiratory function and overall quality of life. These complications, if left undetected, can lead to delayed interventions, resulting in worsening of symptoms, respiratory failure, and in some cases, the need for further invasive procedures. The ability to promptly identify these risks is vital for ensuring timely clinical intervention and improving patient outcomes.

For postoperative recurrent NPC patients, the primary objective is not to segment residual tumor tissue but to delineate vital anatomical structures around the airway that may be at risk due to previous surgical treatment. Precise segmentation of critical airway-related structures—including the larynx, pharyngeal constrictors, and adjacent airway regions—is essential for the identification of potential complications. Even minor inaccuracies in the segmentation of these delicate regions could lead to misdiagnosis or underestimation of postoperative airway risks. For instance, undetected airway narrowing could result in delayed intervention, leading to serious complications such as airway obstruction, difficulty in breathing, or the need for surgical revisions. Moreover, inaccuracies in airway segmentation may also affect the planning and evaluation of post-surgical therapies, including radiation or mechanical ventilation.

TopoWMamba’s architecture effectively addresses these challenges by integrating wavelet-based multi-scale feature extraction with efficient state-space sequence modeling, alongside topology-aware feature extraction. The Wavelet-based Mamba Block (WMB) ensures that both high-frequency details, such as sharp anatomical boundaries, and low-frequency context, like overall structure shapes, are accurately captured. This approach is crucial for delineating complex and subtle structures in the postoperative context, where anatomical changes due to surgery may alter the appearance of critical structures. Additionally, the topology-aware Snake Conv VSS (SCVSS) block enhances boundary delineation by adaptively reordering feature patches, ensuring that the anatomical continuity and spatial relationships of critical structures are maintained, even in the presence of postoperative alterations. The accurate segmentation provided by TopoWMamba can significantly reduce the risk of misdiagnosis, leading to earlier and more effective interventions for airway-related complications.

In summary, TopoWMamba establishes a solid foundation for the postoperative management of recurrent NPC patients by providing anatomically accurate, topologically consistent segmentations of airway-related structures. 

% By minimizing segmentation errors and preserving essential anatomical structures, TopoWMamba holds the potential to improve patient safety, optimize clinical decision-making, and reduce the risk of postoperative airway complications. The precise identification of critical airway structures can help clinicians better monitor and adjust therapeutic interventions, reducing the likelihood of severe complications and improving patient recovery. These advancements highlight the crucial role that automated segmentation models, like TopoWMamba, can play in enhancing clinical workflows and ultimately contributing to better patient outcomes in the management of recurrent NPC.

% \subsection*{CRediT authorship contribution statement}
% Haishan Huang, Pengchen Liang, Jianguo Chen, Naier Lin, Bin Pu: Method design, Conduct experiments, Validation. Haishan Huang, Pengchen Liang, Qing Chang, Luxi Wang: Medical data collection, Writing, Language polishing. Guo Ran, Xia Shen: Writing – review \& editing.

\subsection*{Declaration of competing interest}

The authors state that they have no conflicts of interest related to the creation and publication of this article.

\subsection*{Acknowledgement}
This work is partially funded by the National Natural Science Foundation of China under Grants 6 2002110 and 62372486, and the Natural Science Foundation of Guangdong Province under Grant 2023A1515011179.

\subsection*{Data availability}
Data will be made available on request.

\bibliographystyle{elsarticle-num-names}
\bibliography{reference}

\end{document}